\documentclass[article]{aa}
\usepackage{graphicx}

\newcommand{\etal} {et al.\ }
\newcommand{\be}{\begin{equation}}
\newcommand{\ee}{\end{equation}}
\newcommand{\beq}{\begin{eqnarray}}
\newcommand{\eeq}{\end{eqnarray}}

\newcommand{\aeta}[3]{   #1, A\&A,  #2, #3}

\newcommand{\apj}[3]{    #1, ApJ, #2, #3}

\newcommand{\sph}[3]{    #1, Solar Phys., #2, #3}
\newcommand{\sphp}[1]{   #1, Solar Phys., in press}

\begin{document}

\title{Magnetic Twist and Writhe of Active Regions}
\subtitle{On the Origin of Deformed Flux Tubes}

\author{M.C. L\'opez Fuentes\inst{1}
     \and P. D\'emoulin\inst{2}
     \and C.H. Mandrini\inst{1}
     \and A.A. Pevtsov\inst{3}
     \and L. van Driel-Gesztelyi\inst{2,4,5,6}
        }

\titlerunning{Magnetic Twist and Writhe of ARs}
\authorrunning{L\'opez Fuentes et al.}

\offprints{M. L\'opez Fuentes}

\institute{Instituto de Astronom\'{\i}a y F\'{\i}sica del Espacio, CC. 67,
                  suc. 28, 1428 Buenos Aires, Argentina\\
                 \email{lopezf@iafe.uba.ar \thanks{Fellow of CONICET},
                        mandrini@iafe.uba.ar
                        \thanks{Member of the Carrera del Investigador
                         Cient\'\i fico, CONICET}}
            \and
                Observatoire de Paris, LESIA, FRE 2461 (CNRS), F-92195,
                Meudon, France\\
                \email{pascal.demoulin@obspm.fr}
            \and
                National Solar Observatory, Sacramento Peak, Sunspot, 
                NM 88349, U.S.A.\\  \email{apevtsov@nso.edu}
            \and
                Centre for Plasma Astrophysics, K.U. Leuven,
                Celestijnenlaan 200B, 3001 Leuven, Belgium
             \and
                Mullard Space Science Laboratory, University College London, 
                Holmbury St. Mary, Dorking, Surrey RH5 6NT, UK
             \and
                Konkoly Observatory, Budapest, Pf. 67, H-1525, Hungary }

\date{Received 18 July 2002}

      \abstract{ We study the long term evolution of a set of
22 bipolar active regions (ARs) in which the main photospheric
polarities are seen to rotate one around the other during several
solar rotations.  We first show that differential rotation is not
at the origin of this large change in the tilt angle.  
%We interpret this
%peculiar evolution as being the result of the emergence of
%magnetic flux tubes which are distorted with respect to the
%classical planar $\Omega$-loop shape; then, these flux tubes have 
%a non-null writhe. 
A possible origin of this
distortion is the nonlinear development of a kink-instability 
at the base of the convective zone; this would imply the 
formation of a non-planar flux tube which, while emerging across 
the photosphere, would show a rotation of its photospheric 
polarities as observed.  A
characteristic of the flux tubes deformed by this mechanism is
that their magnetic twist and writhe should have the same sign.
   From the observed evolution of the tilt of the bipoles, we
derive the sign of the writhe of the flux tube forming each AR; while
we compute the sign of the twist from transverse field
measurements. Comparing the handedness of the magnetic twist and
writhe, we find that the presence of kink-unstable flux tubes is
coherent with no more than 35\% of the 20 cases for which the sign
of the twist can be unambiguously determined. Since at most only a
fraction of the tilt evolution can be explained by this process,
we discuss the role that other mechanisms may play in the inferred
deformation. We find that 36\% of the 22 cases may result from the
action of the Coriolis force as the flux tube travels through the
convection zone. Furthermore, because several bipoles overpass in
their rotation the mean toroidal (East-West) direction or rotate
away from it, we propose that a possible explanation for the
deformation of all these flux tubes may lie in the interaction
with large-scale vortical motions of the plasma in the convection
zone, including also photospheric or shallow sub-photospheric 
large scale flows.
}

\maketitle
%%%%%%%%%%%%%%%%%%%%%%%%%%%%%%%%%%%%%%%%%%%%%%%%%%%%%%%%%%%%%%%%%%%%%%%%%%%
\section{Introduction}
\label{introduction}

     Active regions (ARs)  have long been thought to be the manifestation
of the emergence of buoyant magnetic flux tubes having the typical
shape of the letter $\Omega$ (Zwaan \cite{Zwaan85} and references
therein).  These tubes are supposed to be formed at the base of the
convection zone (CZ) from the global toroidal component of the solar
magnetic field (Parker \cite{Parker93}).  After ascending through the
CZ and emerging, their cross-sections with the photosphere are
observed as magnetic field concentrations forming ARs.  The
orientation of positive and negative polarities of ARs with respect to
the East-West direction is opposite in both hemispheres and reverses
from one solar cycle to the next (see Babcock \cite{Babcock61}),
obeying the so called Hale-Nicholson law (Hale \& Nicholson
\cite{Hale25}).  Another fact observed is the tendency of the
preceding polarity to lie closer to the solar equator than the
following one, known as Joy's law (see Hale \etal \cite{Hale19}), the
name of this relationship can be attributed to Zirin (\cite{Zirin88}, p. 307).

\begin{figure*}   %%%%%%%%%%%%%%%%%%%%%%%%%FIGURE 1
      \centering
     \hspace{0cm}
\includegraphics[width=17.cm]{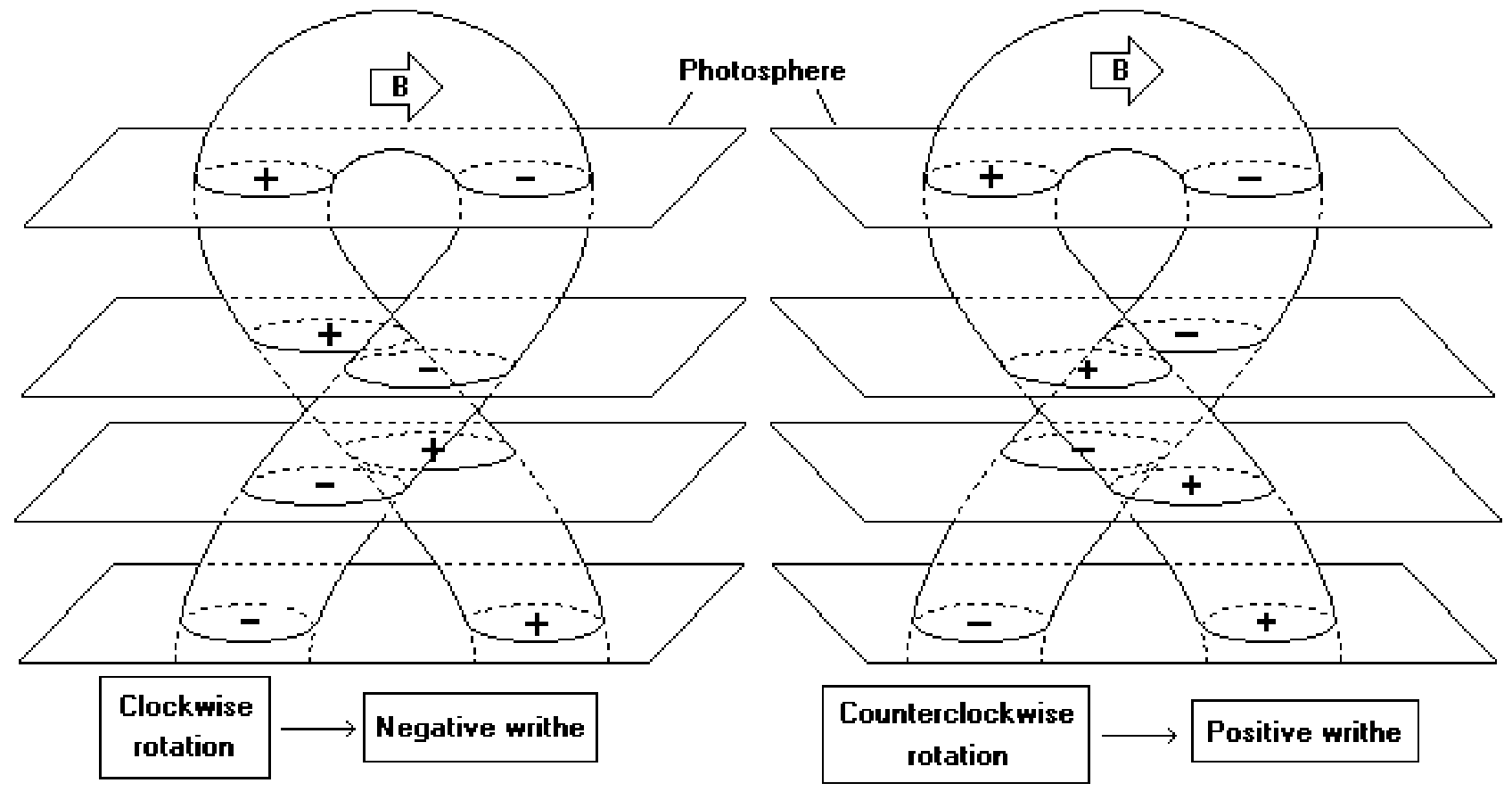}
         \caption{Magnetic flux tubes distorted from the typical
$\Omega$-loop shape.  The sketch in the left panel shows an almost
rigid flux tube with negative writhe; then, as it emerges across the 
photosphere
(horizontal planes) the orientation of the bipolar flux concentrations
changes (they rotate) clockwise.  Correspondingly, in the right panel,
a flux tube with positive writhe will appear as a bipole rotating
counterclockwise.
                 }
            \label{tubes}
\end{figure*}

     However, several ARs do not follow the Hale-Nicholson law and/or
the Joy's law, but their magnetic polarities show proper motions that
are not compatible with the emergence of a planar flux tube
(Cannon \& Marquette \cite{Cannon91}, Tanaka \cite{Tanaka91}, Lites
\etal \cite{Lites95}, Leka \etal \cite{Leka96}, Pevtsov \& Longcope
\cite{Pevtsov98b}).  The evolution of these peculiar ARs has often
been associated with the emergence of distorted magnetic flux tubes.
In particular, there are examples of deformed flux tubes that
are suspected to be formed by the development of a kink instability
(Linton \etal \cite{Linton98}).

     In a recent paper (L\'opez Fuentes \etal \cite{Lopezf00}),
we have analyzed the evolution of a bipolar AR (NOAA 7912) in
which the main polarities were observed to turn one around the
other during several solar rotations.  AR 7912 appeared on the
southern hemisphere as a non-Hale region for cycle 22 in October
1995, becoming a Hale region three solar rotations later (January
1996) after the main polarities were seen to turn by more than
180 degrees. After discarding photospheric or 
shallow sub-photospheric flows that are unlikely to explain why the AR 
was initially formed with a non-Hale orientation, we
interpreted this evolution as the emergence
of a very distorted magnetic flux tube in which the sign of the
twist (a measure of the turning of the field lines around the
axis of the flux tube) was different from the sign of the writhe
(a measure of the spatial turning of the axis of the flux tube).
In a flux tube where the kink instability has developed the sign
of the twist and the writhe should be the same because, as the
instability grows, part of the twist is transferred into writhe
(e.g.  Linton \etal \cite{Linton99}).  Then, we concluded that a
kink instability could not be at the origin of the tube
deformation.  We proposed that it was due to the interaction of
the flux tube with surrounding plasma motions during its emergence
through the CZ.

    %PARAGRAPH Main objectives \\
     In this work we analyze in detail a set of 22 bipolar flux
concentrations, and their subsequent reappearances on the solar disk,
in which the main polarities are observed to rotate one around the
other. We refer to these ARs as rotating tilt-angle ARs (RTARs).
Each reappearance of a given flux concentration is simply
called an AR, which is named using the first NOAA number found in
SGD as given in Tables~\ref{south} and~\ref{north}.  As in the case
of AR 7912, we interpret their evolution as due to the emergence of
deformed flux tubes.  In order to test the role of the kink
instability to originate the flux-tube deformation, we analyze
magnetic field observations and obtain the signs of the twist and the
writhe.  To determine the sign of the writhe we study the evolution of
the tilt angle of the AR, which is defined as the angle that the axis
joining the flux-weighted centers of the main polarities form with the
solar equator.  We measure this angle in synoptic magnetic maps
obtained by the National Solar Observatory at Kitt Peak (NSO/KP).  
On the other
hand, using the linear force-free field equation ($\vec{\nabla}
\times \vec{B}= \alpha \vec{B}$, where $\alpha$ is a constant), we
compute from vector magnetograms obtained mainly at Mees Solar Observatory
(MSO) the value of the parameter, $\alpha_{\rm best}$, that best fits to the
data.  The sign of $\alpha_{\rm best}$ is then used as a proxy for the
sign of the twist.

    %PARAGRAPH Road map \\
     In Section~\ref{writhe}, we describe the data used, together with
the procedure and criteria followed to determine the writhe of ARs.
We outline briefly the method used to compute the sign of the twist in
Section~\ref{twist}.  A summary and an interpretation of the results
is presented in Section~\ref{origin}.  We discuss the relevance of the
kink instability as the origin of the deformation of the flux tubes,
and we analyze the role that other mechanisms, such as the Coriolis
force and external rotational motions of the plasma in the CZ 
or at/near the photospheric surface, may
have in the observed evolution.  Finally, in Section~\ref{conclusion}
we conclude.

\begin{figure*}   %%%%%%%%%%%%%%%%%%%%%%%%%FIGURE 2
      \centering
     \hspace{0cm}
\includegraphics[width=17.cm]{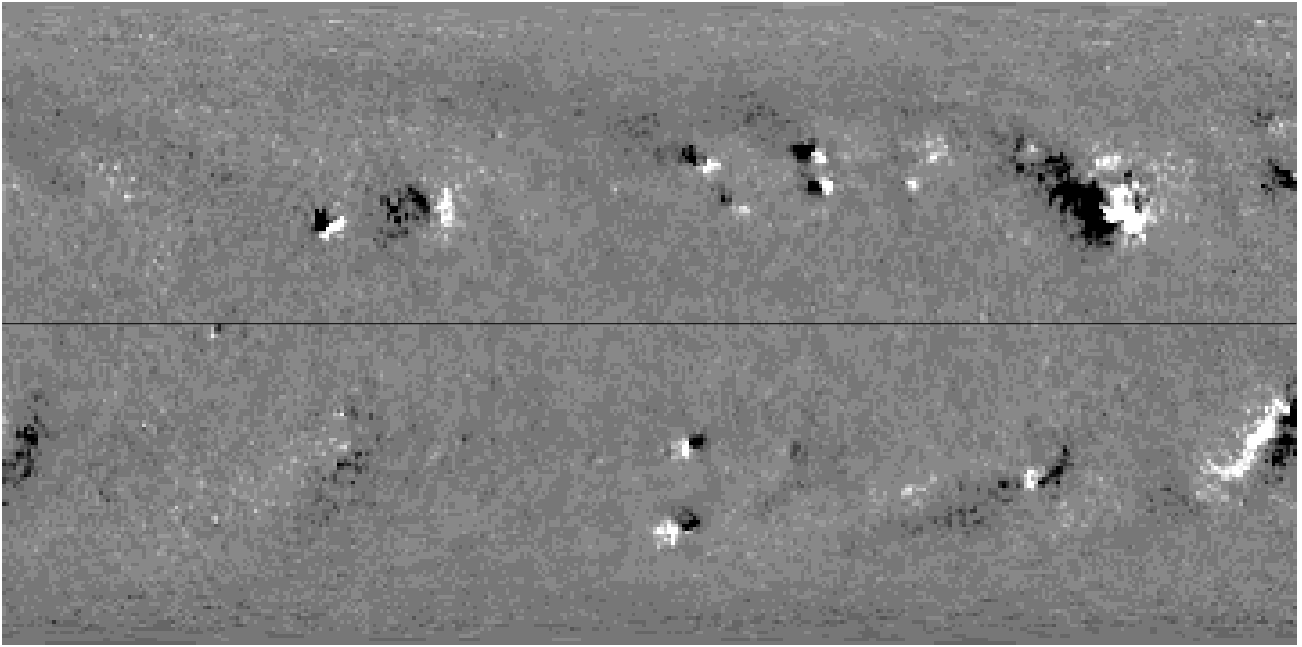}
         \caption{Typical synoptic map from NSO/KP.
White (black) concentrations correspond to positive (negative)
magnetic flux.  Bipolar ARs are easily identified as pairs of flux
concentrations.  The magnetic field has been saturated above
(below) 100 G (-100 G).  This particular map corresponds to
Carrington rotation number 1930 that started on November 28, 1997.
The AR shown in Fig.~\ref{ar_example} is first seen on this map
(top right corner). The horizontal continuous line is the solar
equator.}
            \label{kpmap}
\end{figure*}

%%%%%%%%%%%%%%%%%%%%%%%%%%%%%%%%%%%%%%%%%%%%%%%%%%%%%%%%%%%%%%%%%%%%%%%%%%%
\section{Inferring the writhe of flux tubes from the evolution of
the main polarities}
     %Inferring the writhe of RTARs}
\label{writhe}

\subsection{The relation between the tilt evolution and the writhe}
\label{writhe-tilt}

%     The evolution of ARs at photospheric level can be related to the
%morphology of the flux tubes that give origin to them.  

The observed global evolution of RTARs can a priori be the
consequence of either the drag of plasma flows forcing the magnetic 
flux-tube to rotate or its own internal dynamics. In the first case, 
for example, photospheric or shallow sub-photospheric vortex motions can 
be present where some ARs 
emerge.  After the emergence of a classical planar $\Omega$-loop, the 
drag force applied by such vortex motions could force the main 
polarities to rotate as observed in RTARs. Observations to test such 
hypothesis are not presently available. Another possibility is that 
the $\Omega$-loop was distorted from it planar shape in the CZ itself.  
In such a case the observed rotation of the polarities is a consequence 
of the flux tube morphology as it moves across the photosphere.  In the 
present work we test this second possibility; in particular, the 
possible development of a kink instability in the CZ.  Then, as done
previously (L\'opez Fuentes \etal \cite{Lopezf00}), we can obtain the
sign of the writhe of the tube from the sense of rotation of the main
polarities along the AR life time.  Figure~\ref{tubes} shows two 
almost rigid flux
tubes with their axis distorted in a helix-like curve. The
different planes correspond to consecutive cross-sections of the
flux tubes, giving the evolution of the relative positions of the opposite
polarities as the deformed flux tubes are emerging. As time
goes on, the
magnetic flux concentrations present different tilt angles from
which we can derive the sign of the writhe.  A clockwise rotation of
the polarities indicates a negative (left-handed) writhe, and a
counterclockwise a positive (right-handed) writhe. In these examples
the flux concentrations rotate at photospheric level by 180
deg., but the writhe can be inferred in the same way
for any significant rotation.

\subsection{The data set used}
\label{writhe-data}

     We have used NSO/KP synoptic maps
(see ftp://\-argo.tuc.noao.edu\-/kpvt\-/synoptic\-/README)
to measure systematically the tilt angle of ARs.  The maps consist of
rectangular arrays built from daily full-disk magnetograms using a
triangular weighted integration.  The result of this procedure is
arranged in an array from right to left (West to East) as time
evolves.  The arrays have a 360 by 180 pixels format; then, each pixel
in the horizontal direction corresponds to a heliographic longitude
degree, while in the vertical direction they correspond to the sinus
of the latitude in heliographic coordinates.  Each pixel contains the
time and space averaged magnetic flux per unit surface crossing the
photosphere at a given position on the Sun.  Projection effects are
taken into account by correcting the flux according to the real solar
area covered by a particular pixel and by supposing that the
photospheric field is radial.  Figure~\ref{kpmap} shows one of these
maps as an example, it is possible to identify the ARs as pairs of
black and white concentrations corresponding to negative and positive
magnetic flux, respectively.

\begin{figure*}   %%%%%%%%%%%%%%%%%%%%%%%%%FIGURE 3
\centering
     \hspace{0cm}
\includegraphics[bb=40 255 565 610,height=12cm]{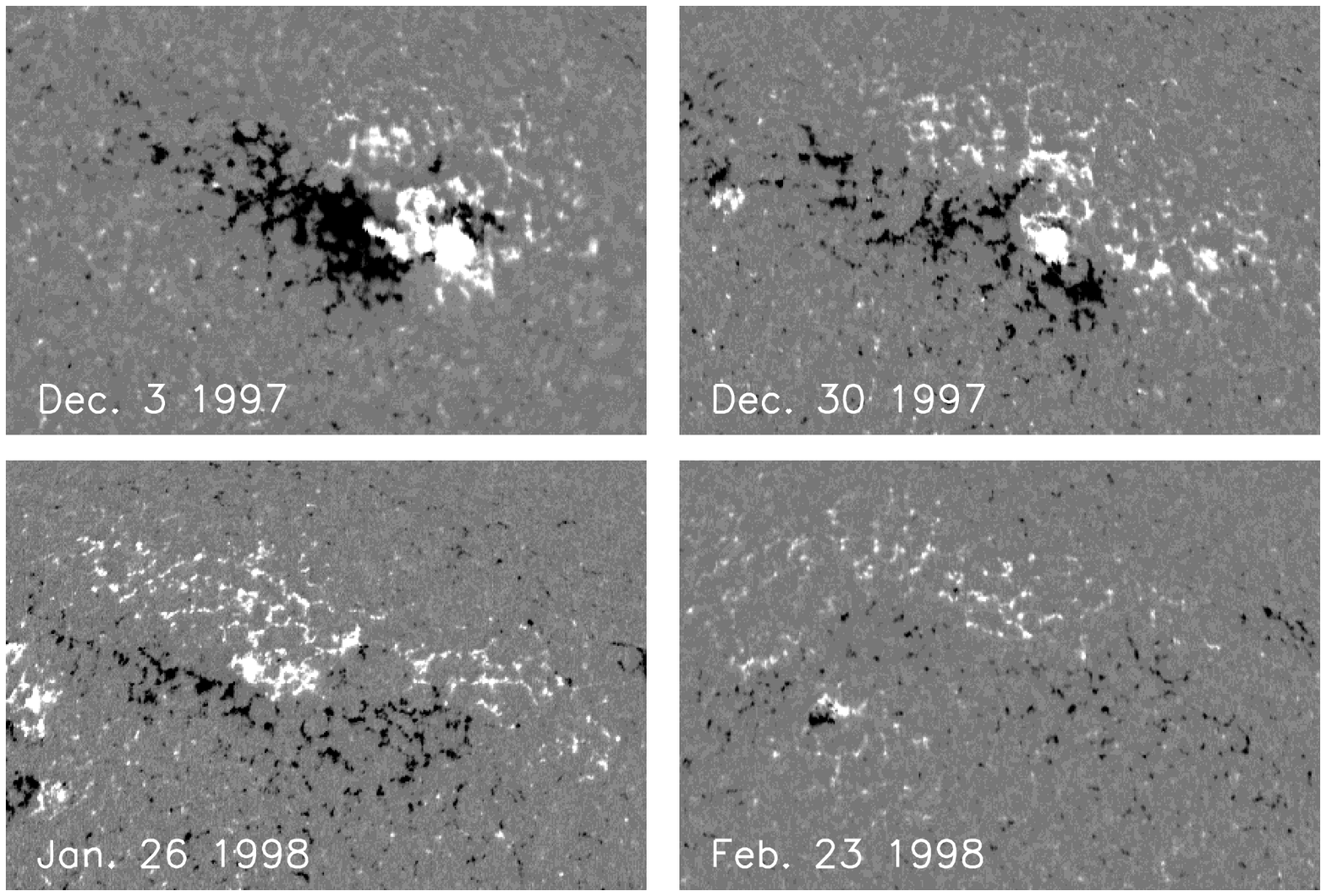}
         \caption{Evolution of AR 8113 from December 1997 to February 1998.
The December 3 magnetogram corresponds to Carrington rotation 1930
shown in Fig.~\ref{kpmap}.  These images come from Kitt Peak full disk
magnetograms obtained on the dates appearing on the panels.  Note the
change in the tilt angle along the four solar rotations (see also
Fig.~\ref{ars1} right panel).  The flux is saturated above (below) 100
(-100) G.  Each panel covers an area of $450 \times 300 \arcsec$.
North is up and West is to the right.}
           \label{ar_example}
      \end{figure*}

\subsection{Description of the procedure}
\label{writhe-procedure}

     We have developed a routine that allows us to compare successive synoptic
maps, and to select from them ARs that survive more than one solar
rotation by visual inspection.  In particular, we are interested in
those ARs where a systematic evolution of the tilt angle is observed.
An example of such regions (AR 8113) is shown in
Figure~\ref{ar_example}, where the variation of the tilt angle of the
bipole is clearly seen. A strong dispersion of the flux is also
evident in the last two solar rotations.  Once a flux concentration is
selected, we isolate from the synoptic map a submatrix including it.
We compute from the data in this submatrix the flux weighted mean
longitude ($\phi$) and latitude ($\lambda$) of the positive (P) and
negative (N) magnetic polarities:
     \begin{equation}
     \phi_{P,N} = \frac{\sum \phi |B|}{\sum |B|},~~~~~~
     \lambda_{P,N} = \frac{\sum \lambda |B|}{\sum |B|}\,,
     \label{pol_position}
     \end{equation}
where $B$ is the magnetic flux crossing the 
photosphere associated to each pixel.  The summation is done over the 
pixels where $|B|$ is above a given value $B_{min}$, for which we have 
chosen 10 G (higher
values, e.g.  $100~G$, give similar results for the first appearance
of an AR, but they do not allow us to follow its decaying phase long
enough).  From the previous parameters, we compute the mean longitude
and latitude, $\bar{\phi} = (\phi_{P}+\phi_{N})/2$ and $\bar{\lambda}
= (\lambda_{P}+\lambda_{N})/2$.  The dipolar size, $S$, and the
tilt angle, $\varphi$, of the AR are defined as:
     \beq
      S &=& R_{\sun} \sqrt{(\phi_P - \phi_N)^2 \cos^2\bar{\lambda}
                            + (\lambda_P -\lambda_N)^2}  \,,
                                                          \label{ar_size} \\
      \varphi &=& \arctan \left(
                         \frac{\lambda_P - \lambda_N}
                         { (\phi_P - \phi_N) \cos \bar{\lambda} }
                               \right) \,,
                                                          \label{ar_tilt}
      \eeq
where $R_{\sun}$ is the solar radius (and both longitude and latitude are
in radians).

    We have also quantified the dispersion of the polarities measuring the
flux weighted mean size of both, positive and negative, flux
concentrations as:
     \begin{equation}
     R_{P,N}=R_{\sun} \frac{\sum
                        \sqrt{(\phi - \phi_{P,N})^2 \cos^2\bar{\lambda}
                            + (\lambda - \lambda_{P,N})^2}|B|}
                        {\sum |B|} \,.  
     \label{pol_size}
     \end{equation}

     The quantities given by Eqs.~(\ref{pol_position}-\ref{pol_size}) are
computed for all the successive appearances of the selected ARs.  We
have also computed the total magnetic flux by simply adding the values
of the flux in the pixels where the field is above $B_{min}$.  These
values are kept in a data base that covers two and a half solar
cycles.

    %PARAGRAPH  The data used in this paper \\
     The present study is limited to a set of 22 ARs for which we can derive
both the sign of the writhe and the $\alpha_{\rm best}$.  
The main results are
summarized in Tables~\ref{south} and~\ref{north}.  The ARs are
grouped according to their reappearances on the solar disk along
successive solar rotations.  We have obtained the NOAA numbers
from the Greenwich Observatory archives (see e.g.
http://\-science.nasa.gov\-/ssl\-/pad\-/solar\-/greenwich.htm),
comparing the heliographic coordinates of the identified ARs to
the positions of sunspots provided in the archives.

\subsection{The criteria to identify reappearances of flux concentrations}
\label{writhe-criteria}

     %PARAGRAPH Intro \& summary of the identification \\
    To test that the supposed reappearances of a flux concentration
indeed correspond to the same AR observed in successive solar
rotations, we have applied certain criteria that we describe in this
section.  First of all, the longitude and latitude of an AR cannot
differ by more than a few heliographic degrees when differential
rotation is taken into account.  On the other hand, we also require
that the evolution of the magnetic flux follows what is expected in
the case of a flux concentration that is emerging, growing and
dispersing afterwards.  These criteria have been implemented in a code
to filter the data set built by visual inspection.

     %PARAGRAPH diff. rotation \\
     Because of the effect of differential rotation, ARs that lie at higher
latitudes appear progressively shifted to decreasing longitudes.
This effect was corrected using the classical expression for differential
rotation ($\omega (\lambda)$):
         \be
         \label{Eq-diff-rot}
         \omega (\lambda)  =  a + b \sin ^{2} \lambda + c \sin ^{4} \lambda \,.
         \ee
     We have taken  $a=14.38 \deg$/day,  $b=-1.95 \deg$/day
and $c=-2.17 \deg$/day,
as given by the cross-correlation analysis of Kitt Peak magnetograms
from 1975 to 1991 (Komm et al.  \cite{Komm93a}).  These values are
very close to the recent ones deduced from seismology measurements
using MDI (e.g.  Charbonneau et al.  \cite{Charbonneau99} and
references therein).  Because the values of the published coefficients
($a,b,c$) are very close, the association of magnetic bipoles along
successive Carrington rotations is independent on the precise
coefficient choice.

     %PARAGRAPH Pratical limits in long. and lat. \\
     We consider that the position of an AR in a given rotation cannot 
differ by more than 5 heliographic degrees in latitude and 8 degrees in
longitude from the previous rotation, to be identified as the same
flux tube.  With this range of confidence, we are taking into account
that the position of an AR can change from one solar rotation to the
next due only to changes in the photospheric distribution of the flux.
This interval is small enough to eliminate confusion with surrounding
ARs and, at the same time, is large enough to include the influence of
the meridional drift of magnetic concentrations, as discussed
below.

     %PARAGRAPH meridional drift \\
     Our measurements show the meridional drift of ARs during their
evolution.
The mean meridional velocity (latitudinal displacement divided by
time averaged along several rotations) is within $\approx
1.8$~m~s$^{-1}$ to $\approx 8.1$~m~s$^{-1}$ (giving a shift in the range 
[$0.3^0,1.7^0$] for one Carrington rotation).  The average value for
all the ARs is $\approx 5.5 \pm 2$~m~s$^{-1}$, this value is coherent
with the one found by Komm et al.  (\cite{Komm93b}) from the study of
Kitt Peak magnetic maps.

     %PARAGRAPH Constraint on the radius and the flux \\
     Concerning the evolution of the size of the polarities in
an AR, it has been observed that as the flux disperses the size increases
with time (Harvey \cite{Harvey93}, van~Driel-Gesztelyi
\cite{vanDriel98}).  In this aspect, our criterion is that $R_{P,N}$
cannot decrease from one solar rotation to the next for a given
concentration to be considered as the reappearance of the same flux
system.  We have also imposed that the magnetic flux of the AR can
have only one maximum.  Using these criteria, we try to remove from
our set cases of nearby and successive emergences of magnetic flux
that form a complex of activity (Gaizauskas et al.  \cite{Gaizauskas83}).
Such multiple emergences can result in a change in the tilt that is
not related to the writhe of the individual flux tubes.  Since these
tests are not enough to fully eliminate these multiple emergences, we
further inspect the magnetic evolution of each selected AR to verify
that the change in the tilt was not critically influenced by them.

     %PARAGRAPH Subset of RTARs analyzed here \\
    After applying the just discussed criteria to the original data set,
we identify a total number $\approx 300$ RTARs, from June 1975 to
December 2000.  We will analyze here only those cases for which we can
determine both the sign of the writhe and of the twist.  A
statistical study of the complete set and its relation to solar
activity will be included in a future paper (L\'opez Fuentes et al.,
in preparation), preliminary results of this work have been discussed
in Mandrini \etal (\cite{Mandrini02}).

\begin{figure}   %%%%%%%%%%%%%%%%%%%%%%%%%FIGURE 4
      \centering
     \hspace{0cm}
\includegraphics[bb=50 180 290 680,width=8.5cm]{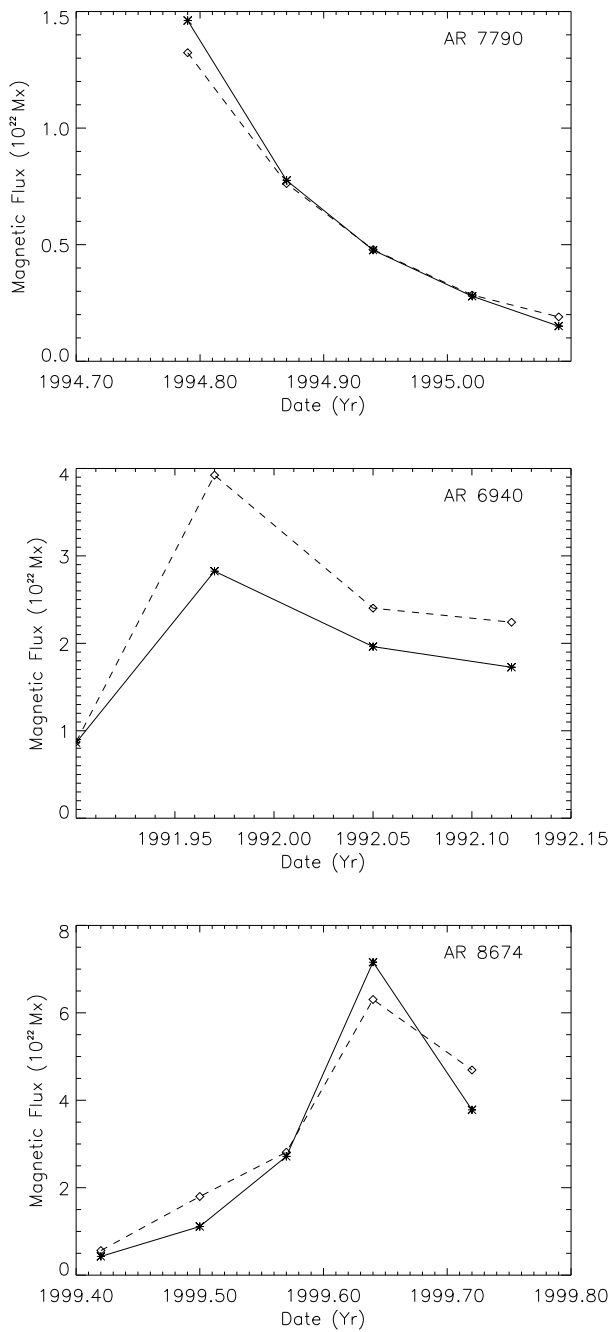}
         \caption{Three different examples of the evolution of the flux
in ARs.  The top panel corresponds to an AR that we start
observing during its decaying phase.  The middle panel shows the
most typical case, an AR with an initial phase of increasing flux
and a long lasting decay.  The bottom panel corresponds to the
evolution of a strange case in which the flux increase lasts
several solar rotations, decreasing later on. The continuous line
corresponds to the positive flux and the dashed line to the
negative one.
                 }
            \label{arsflux}
      \end{figure}

\subsection{Evolution of the magnetic flux}
          \label{writhe-flux}

    %PARAGRAPH Not so peculiar from the flux and for Link flux-size AR ! \\
     The analyzed RTARs have no peculiar characteristics in
terms of the evolution of the magnetic flux as compared to the
classical ARs (not or weakly rotating).  The mean flux is of the
order of $10^{22}$ Mx, which is the average value for the group of
strong bipoles forming ARs (Wang \& Sheeley \cite{Wang89}). The
selection of strong bipoles is implicitly associated with the
duration of one Carrington rotation because of their
characteristic decay time. ARs typically decay at a rate of 1 to 2
$ \times 10^{20}$~Mx/day (Schrijver \& Zwaan \cite{Schrijver00}
p.~114-115); then, small ARs of $5 \times 10^{21}$~Mx survive 
for about 25-50 days, while the smaller ones are excluded. We
have also found that the magnetic flux is higher for ARs where $S$
(dipolar size) is larger, in agreement with previous results (e.g.
Wang \& Sheeley \cite{Wang89}; Howard \cite{Howard92}).

    %PARAGRAPH Three kind of flux evolution \\
     Figure~\ref{arsflux} shows three ARs representing different cases of
flux evolution.  The first example (Fig.~\ref{arsflux} top panel)
corresponds to AR 7790 in which the magnetic flux decreases
continuously.  This means that this AR has been identified after its 
emergence phase has ended, and so we observe it along its
decaying phase (5 solar rotations).  The second case shown in
Figure~\ref{arsflux} (middle panel) is AR 6940 in which the increase
of flux lasts just from the first to the second solar rotation, then
this AR has probably been selected at the beginning of its emergence.
The bottom panel corresponds to AR 8674 in which the flux increases
during most of its lifetime, the decaying phase starts as late as the
fourth solar rotation in which it is observed.

    %PARAGRAPH Generality on the flux evolution, then concl. for RTARs \\
     In general, the magnetic flux of ARs evolves in such a way that it shows
an initial rapid increase (in a time scale of weeks), followed by a
long lasting decay (in a time scale of months, see Harvey
\cite{Harvey93}, van~Driel-Gesztelyi~\cite{vanDriel98}).
Nevertheless, we have observed cases in which the flux increase lasts
as long as 3 solar rotations, much more than what is usually believed
(see previous references).  This long-term flux increase is observed
only in a small number of the RTARs, 3 over 22 regions (AR 8674
in Fig.~\ref{arsflux} shows the longest period of flux increase, while in
the other two cases the flux increases only from the first to the third
rotation).  We may ask ourselves if this behavior is inherent to the
RTARs but, since very little is known about their long term
evolution, a further investigation based on a larger data set is
needed (L\'opez Fuentes et al., in preparation).

\subsection{Tilt of active regions}
          \label{writhe-tilt-AR}

    %PARAGRAPH  Conventions for the tilt \\
    The evolution of the tilt angle for
8 ARs is shown in Figures~\ref{ars1} and
\ref{ars2} for the northern and southern hemisphere, respectively.
The mean orientation of the positive and negative polarities of ARs with
respect to the East-West direction is opposite in both hemispheres,
and reverses from one solar cycle to the next (Hale-Nicholson law).
The observed orientation is the one shown in Figures~\ref{ars1} and
\ref{ars2}, but, when discussing the plausible mechanisms originating the
variation in the tilt, it is
better to remove these hemispheric and cycle dependences.  More
precisely, we set in Tables~\ref{south} and~\ref{north} the origin
of the tilt angle in the East-West direction
following the Hale-Nicholson
law.  With this convention, the relaxation, or not, to the East-West
direction is easily seen (Section~\ref{origin-Coriolis}).

    %PARAGRAPH Errors on the tilt \\
    The errors in the tilt are computed  combining the independent
errors coming from each pixel. In practice, this is done by
propagating the error of each parameter (position and field
strength) on which the tilt depends to first order in
Eq.~\ref{ar_tilt}, and taking afterwards the square root of the
summation of the terms in the propagation to the second
power. We assume that the error in the position is half a
pixel of the synoptic map and that the error per pixel in the
magnetic field is 5~G. This last value is a large upper bound since
the flux in each pixel of a synoptic map is computed averaging
several individual magnetograms for which the errors are of
$\approx 7$~G per pixel, this implies an error of about 0.5~G per pixel
in the synoptic map (using only the classical 
propagation of errors).  
However, there are certain systematic
errors in the original data used to build up the synoptic maps,
such as a deficit of flux in strong field regions that we cannot
take into account (Harvey J. private communication). We have found
that, when taking 5~G for the field error, the contribution due to
the error in the flux is in general of the same order as the
contribution due to the error in the position.

\subsection{The contribution of differential rotation to the tilt change}
       \label{writhe-diff.rot}

\begin{figure*}   %%%%%%%%%%%%%%%%%%%%%%%%%FIGURE 5
\centering
\hspace{0cm}
\includegraphics[bb=40 170 580 690,width=17cm]{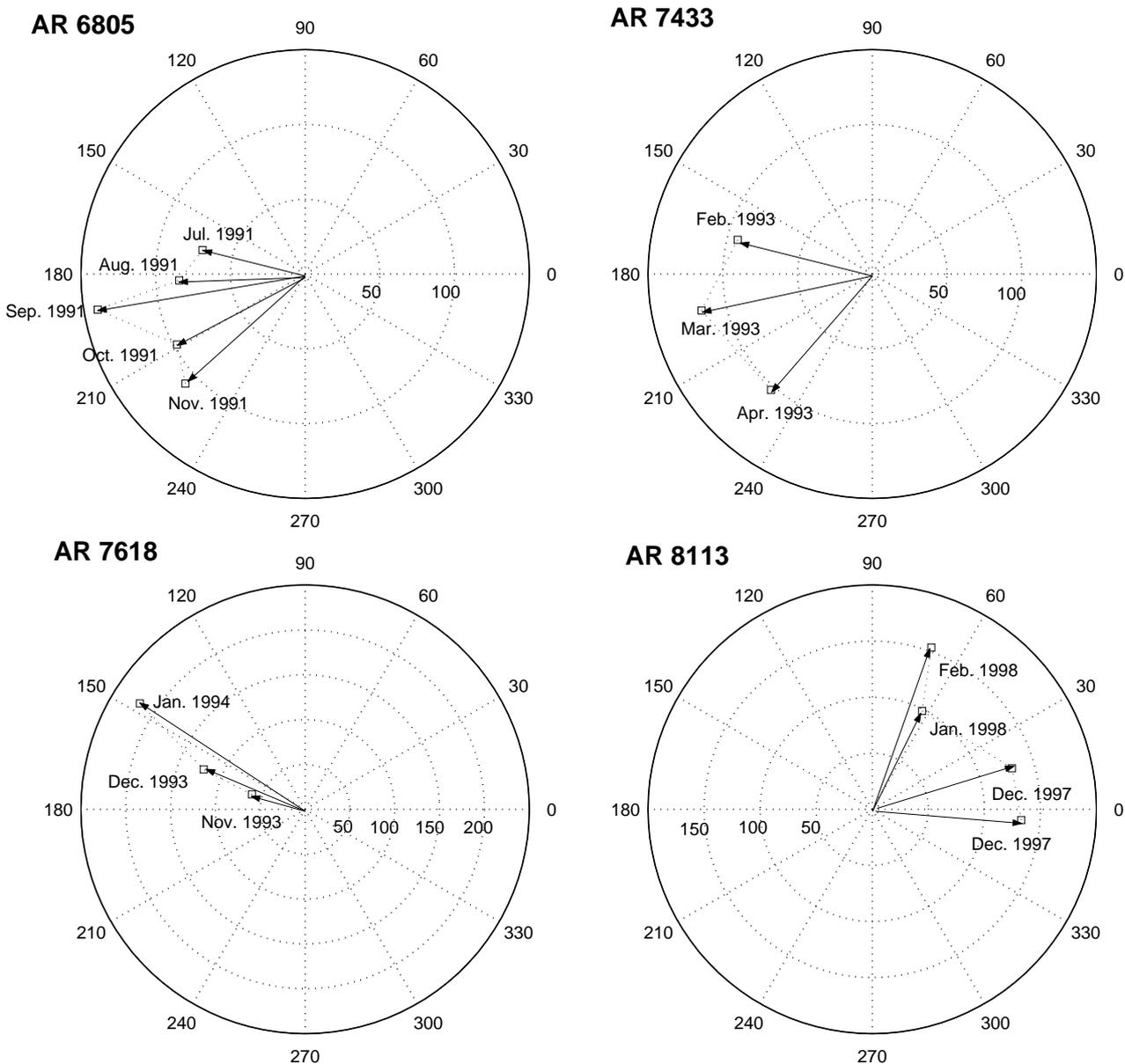}
         \caption{Four RTARs observed on the North hemisphere 
between 1986 and 1998.
The center of the polar plots corresponds to the position of the
center of the negative polarity, while the arrows and squares
indicate the relative position of the positive polarity.  The tilts
are shown as measured on the Sun, except that they are
corrected for differential rotation (Section~\ref{writhe-diff.rot}).
The AR NOAA numbers appear in the upper left corner of the panels and
the labels indicate the dates of the successive reappearances of the
ARs.  Three of these ARs (AR 6805, AR 7433, AR 8113) rotate in the
counterclockwise (CCW) direction and one (AR 7618) in the clockwise
(CW) one.}
            \label{ars1}
      \end{figure*}

\begin{figure*}   %%%%%%%%%%%%%%%%%%%%%%%%%FIGURE 6
      \centering
     \hspace{0cm}
\includegraphics[bb=40 170 580 690,width=17cm]{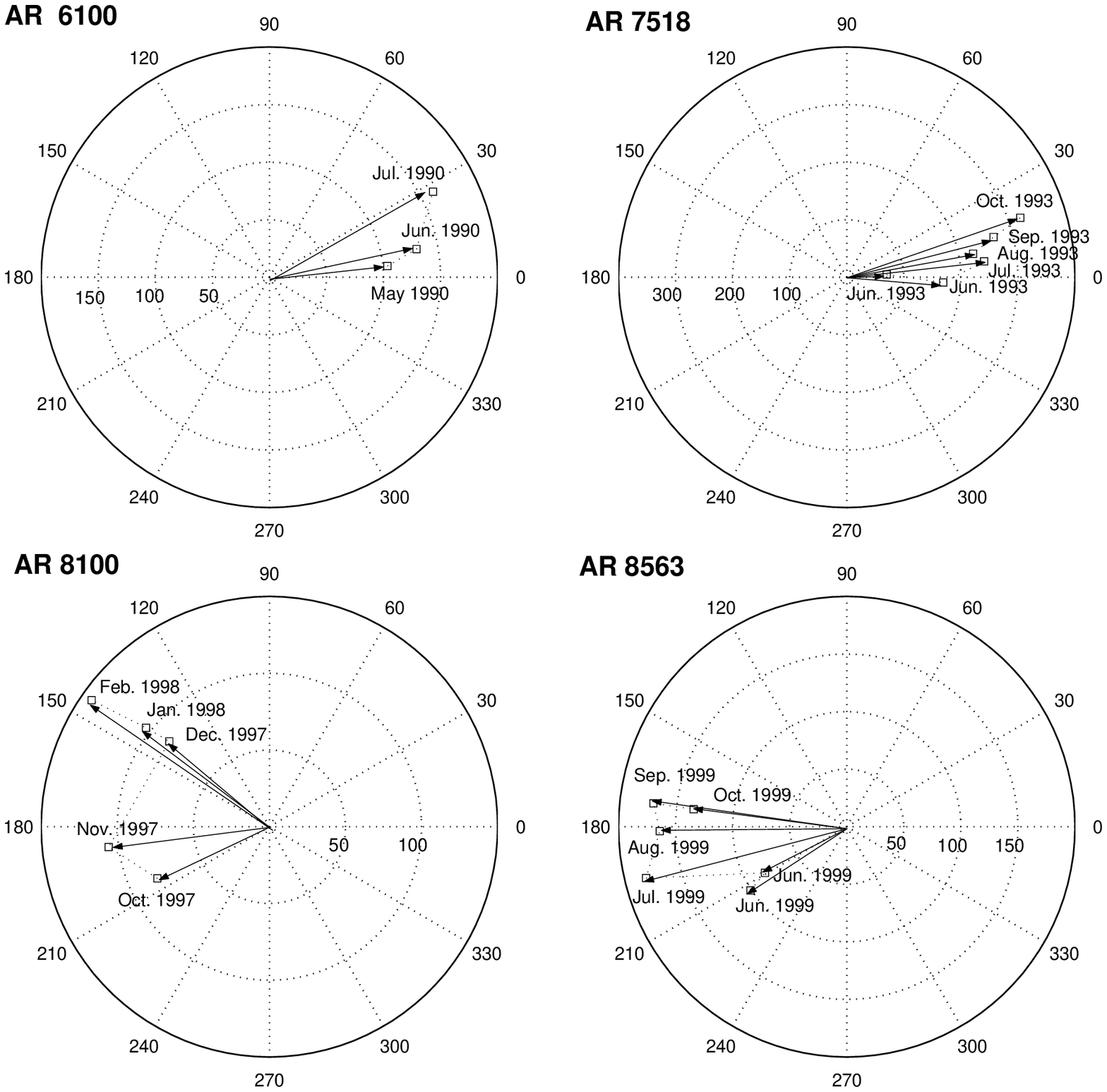}
         \caption{Idem Figure~\ref{ars1} for the South hemisphere.
Two of these ARs (AR 6100, AR 7518) rotate in the CCW direction and
two (AR 8100, AR8563) in the CW one. The tilt angles are corrected for 
differential rotation.}
            \label{ars2}
      \end{figure*}

    %PARAGRAPH  Correction from diff.rot. of the tilt \\
      Differential rotation certainly contributes to modify the
tilt angle of ARs.  The effect of differential rotation is maximum
when the drag force at the photospheric surface becomes so efficient
that it couples the magnetic flux tube to the surrounding plasma.
To subtract the contribution of differential rotation at the
photosphere from Eq.~\ref{ar_tilt}, we have divided each Carrington
rotation in several temporal steps to make our computation more 
precise.  The variation of the tilt angle
corrected in this way is illustrated in Figures~\ref{ars1} and
\ref{ars2} for several ARs from the North and South hemispheres.
The fifth and sixth columns of Tables~\ref{south} and~\ref{north}
show the total rotation angle (final minus initial tilt angle)
computed with and without the effect of differential rotation. The
errors in $\Delta \varphi$ and $\Delta \varphi_{\rm cor.}$,
computed as described in Section~\ref{writhe-tilt-AR}, are almost
the same; that is why, we have included them only for the
$\Delta \varphi_{\rm cor.}$ values. 

     Differential rotation has its most significant effect when
the AR is oriented in the North-South direction, which is rare
according to the Hale-Nicholson law.  In the present data set, we
find that this is the case of AR 4711 and of two ARs with an important
tilt change; AR
8100 that rotates $\approx 150^{0}$ and AR 8113 that
rotates $\approx 140^{0}$ when no correction for the differential
rotation is applied.  These two last examples start as ``normal''
ARs in the sense of the Hale-Nicholson law, but become non-Hale
ARs later on.  The influence of differential rotation is clearly
evident in these two ARs since most of the change in the tilt
angle after their third appearances is provided by it
(Figs.~\ref{ars1} and~\ref{ars2}); however, a significant total
rotation angle is still present ($\approx -60^{0}$ and $\approx
74^{0}$, respectively).

    %PARAGRAPH  Effect of diff.rot. on  the tilt
    The results summarized in Tables~\ref{south} and~\ref{north},
together with polar plots similar to those in Figures~\ref{ars1} and
\ref{ars2} for the full set analyzed here, show that for most of the
ARs $\Delta \varphi$ and $\Delta \varphi_{\rm cor.}$ have comparable
values.  In the few cases discussed above, where differential rotation
has a large effect, a significant rotation of the AR polarities still
remains.  Then, we conclude that differential rotation is not the main
mechanism responsible for the rotation of the bipoles.  In all cases,
the correction of the tilt angle by differential rotation does not
change its sign (which is equivalent to keep the inferred sign of the
writhe unchanged) and, therefore, it does not alter our conclusions.

%%%%%%%%%%%%%%%%%%%%%%%%%%%%%%%%%%%%%%%%%%%%%%%%%%%%%%%%%%%%%%%%%%%%%%%%%%%
\section{Determination of the twist of RTARs}
         \label{twist}

\subsection{The data used}
         \label{twist-data}

     It is not possible to obtain a direct measurement of the twist of the
magnetic flux tubes that form ARs.  Nevertheless, computations of the
global value of the force-free field parameter, $\alpha_{\rm best}$,
provide a proxy for the sign of the magnetic twist in the tubes
forming ARs (see Section \ref{twist-interpret}).  Except for two
regions (AR 4711 and AR 6100), for which we have data from the
magnetograph of the Marshall Space Flight Center (MSFC) (Hagyard
\cite{Hagyard82}), we have used vector magnetograms from the Haleakala
Stokes Polarimeter at MSO (Mickey \cite{Mickey85})
to determine $\alpha_{\rm best}$.  This instrument provides Stokes I,
Q, U and V profiles of the Fe I $\lambda\lambda$ 6301.5, 6302.5
\AA~doublet with a 25 m\AA~pixel$^{-1}$ dispersion.  A typical
magnetogram consists of a 2' x 2' array with a spatial resolution of
less than 3'' and is completed in about one hour.  Each
raster point provides the strength of the magnetic field components
parallel and transverse to the line of sight, as well as the azimuth
of the transverse field (for a detailed description of the data
processing see i.e.  Pevtsov \& Canfield \cite{Pevtsov98a}).  The
azimuth has a 180 degree ambiguity that is resolved by the method
described in Canfield \etal (\cite{Canfield93}).  To avoid projection
effects the magnetograms are transformed to disk-center heliographic
coordinates.

\subsection{The computation of $\alpha_{\rm best}$}
           \label{twist-comp}

      %PARAGRAPH: General method \\
     In force-free magnetic field configurations the rotation of the field
and the magnetic field itself are related by: $\vec \nabla \times \vec B =
\alpha \vec B$. The projection of the force-free equation in the
$z$ direction (being $z$ normal to the photosphere) gives:
     \begin{equation}
     \frac{\partial B_y}{\partial x} - \frac{\partial B_x}{\partial y}
                      = \alpha B_z\,.
     \label{current}
     \end{equation}
%    The right hand side of Eq.~(\ref{current}) corresponds to the
%current density in the $z$ direction. 
In this way maps of $\alpha$
can be created from vector magnetograms. These maps show in general a
large variation of $\alpha$ (both in sign and magnitude) within an AR.
However, there is frequently a given overall sign of $\alpha$
associated to an AR. A single value of $\alpha_{\rm best}$ for the
linear force-free field that best fits the AR magnetic
field is obtained by a least-squares method (Pevtsov et al.
\cite{Pevtsov95}). When more than one magnetogram is available for a
particular AR, we compute the standard deviation that is shown in the
third column in Tables~\ref{south} and~\ref{north} as the error
of $\alpha_{\rm best}$.

\subsection{$\alpha_{\rm best}$ in the selected ARs}
           \label{twist-selectedARs}

      %PARAGRAPH: Coherence of $\alpha_{\rm best}$ with rotations  \\
     When data are available for the successive rotations of the same AR,
we find in general a coherent sign for $\alpha_{\rm best}$.  However,
there are 4 cases where the sign of $\alpha_{\rm best}$ changes along
the evolution of the region; these cases are: AR 6855, AR 8243, AR
7518, AR 8100 (NOAA numbers of the first appearances).  For the first
two cases the change in the sign of $\alpha_{\rm best}$ is significant
(see Table~\ref{north}), so we will not use these ARs in our analysis.
For AR 7518, the change in the sign of $\alpha_{\rm best}$ occurs only
in the last appearance and its error is much larger than its mean
value; therefore, we consider that this measurement is irrelevant.
Finally, the evolution of AR 8100 along five solar rotations has been
studied in detail (Green \etal \cite{Green02}) by us; we have found
(using coronal observations) that, except in the first rotation, the
twist was always positive. Then, we have decided to include AR 7518
and AR 8100 (considering in this case a positive twist) in our study.
For some of the 20 remaining ARs (ARs 7417,
7433, 7618, 7882, 8293, 8674) the value of $\alpha_{\rm best}$ is
largely uncertain; so that, within the error bars, there is a non-null
probability that the twist has the opposite sign than the mean
$\alpha_{\rm best}$.  However, we have decided to keep such cases
because there is still significant information in them.  The
uncertainties in $\alpha_{\rm best}$ are reflected in
Figure~\ref{twist_writhe}.

\subsection{Hemispheric rule}
\label{twist-hemispher}

    %PARAGRAPH Previous work on hemispheric rule \\
     Seehafer (\cite{Seehafer90}) showed that the helicity is predominantly 
negative in the northern hemisphere and positive in the southern one. 
This result was
confirmed and quantified computing $\alpha_{\rm best}$ by Pevtsov
\etal (\cite{Pevtsov95}, \cite{Pevtsov01}), Longcope \etal
(\cite{Longcope98}) and Tian \etal (\cite{Tian01}).  The dominance
of negative (resp. positive) $\alpha_{\rm best}$ in the northern
(resp.  southern) hemisphere is in the range of 62~\% to 75~\%.

    %PARAGRAPH Our work on hemispheric rule \\
     Our sample (see Tables~\ref{south} and~\ref{north}) includes one order of
magnitude less ARs than the previous studies, having a low statistical
significance.  Nevertheless, we see almost the same tendency in the 
RTARs for hemispheric dominance; that is to
say, if we consider the measurements of $\alpha_{\rm best}$ for each
rotation independently, as done in the previous studies, we have a
dominance of 18 negative values of $\alpha_{\rm best}$ over 23 cases
(so 78~\%) for the northern hemisphere, and a dominance of 12 positive
values over 17 cases (so 70~\%) for the southern hemisphere.  If we
count only the magnetic flux tubes (relating successive
rotations, a procedure which takes into account the physics but
decreases the statistics), the dominance turns out to be 10 over 12
cases, so 83~\%, and 6 over 8 so 75~\%, respectively.  The slightly
higher values of the dominance percentages found for the RTARs
are not statistically meaningful to claim for any typical property.
As for the magnetic flux (Section~\ref{writhe-flux}), RTARs are
similar to what we may call ``normal'' ARs with a flux large enough
to survive more than one rotation.

\subsection{Interpretation of $\alpha_{\rm best}$}
\label{twist-interpret}
      It is not straightforward to interpret the meaning of
$\alpha_{\rm best}$ in terms of the original properties of the magnetic
flux tubes in the CZ.  Because magnetic measurements are obtained in a
region of the atmosphere where the magnetic field is still confined by
the plasma, the magnetic field is expected to keep most of its
subphotospheric properties; in particular, $\alpha_{\rm best}$ is a
trace of how the tube was twisted.  But this is not a quantitative
measurement of the twist because magnetograms give information at only
a few cuts of the magnetic tube.  Moreover, the subphotospheric twist
is expected to be progressively transferred to the corona by torsional
Alfv\'en waves (Longcope \& Welsch \cite{Longcope00}).  Keeping this in
mind, we will focus only in the sign of $\alpha_{\rm best}$ as a proxy
for the sign of the magnetic ``twist'' in the flux tube.

%%%%%%%%%%%%%%%%%%%%%%%%%%%%%%%%%%%%%%%%%%%%%%%%%%%%%%%%%%%%%%%%%%%%%%%%%%%
\section{Possible origin of RTARs}
        %section{Results and discussion} \\
     \label{origin}

     The development of the kink instability in the CZ has
been several times invoked to explain the peculiar evolution of
some ARs.  The detailed analysis of a particular RTAR, as well as
the review of other published examples (L\'opez Fuentes \etal
\cite{Lopezf00}), have shown that few cases can be explained as
due to the kink instability.  The main objective of the present
study is first to test this hypothesis on the largest set of data
presently available; that is to say, ARs where the sign of the
writhe and the twist can be both unambiguously determined.  After
doing so, we will explore other possible mechanisms that can be at
the origin of the RTARs.

%___TABLE_1_2___TABLE_1_2______TABLE_1_2______TABLE_1_2__________________
\begin{table}
         \caption[]{List of the RTARs for the South hemisphere.
The first column gives the NOAA number of each appearance for which
$\alpha_{\rm best}$ is available, NOAA numbers that do not belong to
the same magnetic flux tube are separated by a blank line.  The second
column gives the date of the central meridian passage (CMP) of the
corresponding NOAA region.  The following four columns show,
respectively, the value of the force free parameter $\alpha_{\rm
best}$ (in units of $10^{-8}$m$^{-1}$), the tilt angle for the first
rotation $\varphi_{0}$ in degrees (computed as described in
Section~\ref{writhe-tilt-AR}, and taken as $0$ when the bipole is
oriented in the East-West direction following the Hale-Nicholson law),
the total rotation angle of the AR not corrected ($\Delta\varphi$) and
corrected ($\Delta \varphi_{\rm cor.}$) for differential rotation.
The total rotation angles are computed for the full period in which
each AR is observed in the synoptic maps (not just when vector
magnetograms were available). The last column indicates
the way the AR evolves with respect to East-West direction
(a: rotates away from it,
o: rotates towards but overpasses it by more than 10$^{0}$, 
r: relaxes to it within 10$^{0}$, 
t: rotates towards without reaching it within 10$^{0}$) and
the possible deformation mechanism
(C: Coriolis force, K: kink instability, ?: unidentified).
   }
\label{south}
\(
\begin{array}{lc|crrr|rr}
\hline
\noalign{\smallskip}
{\rm AR} &{\rm Date}& {\alpha_{\rm best}} & \varphi_{0} &
\Delta \varphi &\Delta \varphi_{\rm cor.} & \multicolumn{2}{c}{\rm Evol. \&} \\
       &{\rm of~CMP}  &   &  &      &     & \multicolumn{2}{c}{\rm Mech.}\\
\noalign{\smallskip}
\hline
\noalign{\smallskip}
4711 &06/02/86&~-5.0\pm0.5&75.&~-41.&-14.\pm3.& t & K,C   \\
     &        &            &     &    &         &   &     \\
6100 &15/06/90&~~1.4\pm0.3&~5.&~~19.&~22.\pm1.& a & K   \\
     &        &            &     &    &         &   &     \\
6990 &31/12/91&-2.4~~~~~~~&11.&~-20.&-20.\pm2.& r & K,C \\
     &        &            &     &    &         &   &     \\
6940 &28/11/91&~~0.3\pm0.4&28.&~~22.&~22.\pm4.& a & K   \\
6982 &25/12/91&~~5.6\pm1.9&   &     &         &   &     \\
7012 &21/01/92&~~0.7\pm0.5&   &     &         &   &     \\
     &        &            &     &    &         &   &     \\
7518 &05/06/93&~~0.3\pm1.2&~4.&~~15.&~15.\pm2.& a & K   \\
7530 &31/06/93&~~0.4\pm0.9&   &     &         &   &     \\
7553 &28/07/93&~~0.8\pm0.6&   &     &         &   &     \\
7566 &23/08/93&~~1.0\pm0.9&   &     &         &   &     \\
7581 &20/09/93& -1.5\pm7.1&   &     &         &   &     \\
     &        &            &     &    &         &   &     \\
8100 &02/11/97& -1.8\pm1.6&25.&-148.&-60.\pm2.& o & ?   \\
8124 &27/12/97&~~3.6\pm1.0&   &     &         &   &     \\
%8142 &23/01/98&~-0.5~~~~~~&  &     &         &   &     \\
     &        &            &     &    &         &   &     \\
8293 &08/08/98&~~0.5\pm0.8&~8.&~-77.&-39.\pm2.& o & ?   \\
8323 &04/09/98&~~0.2~~~~~~&   &     &         &   &     \\
     &        &            &     &    &         &   &     \\
8674 &20/08/99&~~4.1\pm4.3&31.&~-42.&-40.\pm7.& r & C   \\
\noalign{\smallskip}
\hline
\end{array}
\)
\end{table}

\begin{table}
\caption[]{Idem Table~\ref{north} for the North hemisphere.}
\label{north}
\(
\begin{array}{lc|crrr|rr}
\hline
\noalign{\smallskip}
{\rm AR} &{\rm Date}& {\alpha_{\rm best}} & \varphi_{0} &
\Delta \varphi &\Delta \varphi_{\rm cor.} & \multicolumn{2}{c}{\rm Evol. \&} \\
       &{\rm of~CMP}  &   &  &      &     & \multicolumn{2}{c}{\rm Mech.}\\
\noalign{\smallskip}
\hline
\noalign{\smallskip}
6805 &30/08/91& -2.0\pm 1.1&~-13.&103.&~56.\pm3.& o & ?   \\
     &        &            &     &    &         &   &     \\
6855 &03/10/91& -1.8\pm 1.3&~~22.&-26.&-28.\pm1.& r & ?   \\
6893 &30/10/91&~~1.2\pm 0.4&     &    &         &   &     \\
6936 &26/11/91& -4.1~~~~~~~&     &    &         &   &     \\
     &        &            &     &    &         &   &     \\
7205 &20/06/92&~~4.2\pm 1.1&~-22.&~19.&~13.\pm2.& r & K,C \\
7262 &13/08/92&~~0.5~~~~~~~&     &    &         &         \\
     &        &            &     &    &         &   &     \\
7417 &08/02/93& -0.3\pm 0.3&~~-5.&~10.&~10.\pm1.& r & C   \\
     &        &            &     &    &         &   &     \\
7433 &26/02/93& -0.4\pm 0.6&~-14.&~72.&~63.\pm2.& o & ?   \\
     &        &            &     &    &         &   &     \\
7618 &18/11/93&~~0.2\pm 0.5&~-16.&-13.&-17.\pm2.& a & ?   \\
     &        &            &     &    &         &   &     \\
7645 &01/01/94& -0.6\pm 0.3&~-12.&~~9.&~~~7.\pm1.& r& C   \\
     &        &            &     &    &         &   &     \\
7640 &26/12/93& -0.7\pm 1.3&~~-5.&~12.&~11.\pm1.& r & C   \\
7654 &22/01/94& -1.2\pm 0.6&     &    &         &   &     \\
     &        &            &     &    &         &   &     \\
7762 &08/08/94& -1.6\pm 1.0&-176.&~49.&~46.\pm3.& t & C   \\
7771 &03/09/94& -0.5\pm 0.7&     &    &         &   &     \\
     &        &            &     &    &         &   &     \\
7790 &18/10/94& -0.9\pm 0.4&~-12.&~41.&~37.\pm3.& o & ?   \\
     &        &            &     &    &         &   &     \\
7830 &23/01/95& -0.6\pm 0.4&~-25.&~36.&~35.\pm2.& r & C   \\
     &        &            &     &    &         &   &     \\
7882 &24/06/95& -1.0\pm 1.3&~~-3.& -7.&-9.\pm2.  & a & K   \\
7891 &21/07/95& -24.\pm 45.&     &    &         &   &     \\
%7897 &16/08/95& -8.6\pm 18.5 &  &    &         &   &     \\
7897 &16/08/95& -9.0\pm 18.&     &    &         &   &     \\
     &        &            &     &    &         &   &     \\
8113 &02/12/97& -1.1\pm 0.2&~~-4.&139.&~74.\pm2.& o & ?   \\
8126 &30/12/97& -1.7~~~~~~~&     &    &         &   &     \\
     &        &            &     &    &         &   &     \\
8243 &18/06/98& -2.0\pm 0.7&~~12.&-21.&-25.\pm3.& o & ?   \\
8269 &16/07/98&~~7.8~~~~~~~&     &    &         &   &     \\
\noalign{\smallskip}
\hline
\end{array}
\)
\end{table}

%____TABLE_1_2______TABLE_1_2______TABLE_1_2______TABLE_1_2__________________

\subsection{How do RTARs rotate?}
  \label{relaxation}

%PARAGRAPH: On the rotation sense \\
 The set of RTARs studied here rotate in a coherent way; that
is to say, the sense of rotation is always the same along
their evolution. However, we have found that in 7 cases, after 
correcting for
differential rotation, the general rotation trend changes either 
between the
second and third solar rotations or at the end of the AR evolution
(see e.g. AR 8100 in Fig.~\ref{ars2}). The angle by which the RTARs
rotate back is always of a few degrees, so, of the same order as 
the error
in the total rotation angle ($\Delta \varphi$ or 
$\Delta \varphi_{\rm cor.}$);
therefore, this does not alter the general trend.

%PARAGRAPH: Final direction - cases with relaxation \\
  We can now investigate which is the direction towards which most of
the bipoles relax; that is to say, which is the last observed direction of
the axis joining both polarities. We find that 8 ARs relax to
the East-West direction (where we have set the origin of tilt angle) 
within $10^{0}$; those marked with an $r$
on the left side of the last column in Tables~\ref{south} and~\ref{north}
(this can be easily seen by adding $\varphi_{0}$
to $\Delta \varphi_{\rm cor.}$). Moreover, only 2
ARs (those marked with a $t$ on the left side of the last
columns in Tables~\ref{south} and~\ref{north}) relax towards this
direction without reaching it, but both
are peculiar ARs; AR 4711 starts with a very large tilt angle and
AR 7762 is a non-Hale AR. It is worth noting that these 10 cases
have an evolution similar to the average of the sunspot groups as
found by Howard et al. (\cite{Howard00}, and references therein).
However, our results are based on much longer time scales (several
Carrington rotations) while the relaxation towards the mean tilt
found by Howard et al. occurs on a 3 to 6 days time scale.
Longcope \& Choudhuri (\cite{Longcope02}) have modeled this behavior
as the relaxation of the turbulent perturbations set in the 
magnetic flux tube when it crosses the upper part of the CZ.

%PARAGRAPH: Final direction - cases without relaxation \\
    An important fraction of the ARs (7/22, those marked with an
$o$ on the left side of the last columns in Tables~\ref{south} 
and~\ref{north}) rotates towards
the East-West direction, but after reaching it, they continue
rotating and overpass it by more than $10^{0}$ (with a mean value
of $30^{0}$). This kind of ARs have been found previously by
Cannon \& Marquette (\cite{Cannon91}, their regions 1-4). In a
previous work Marquette \& Martin (\cite{Marquette88}) studied in
detail region 1, that rotates by $\approx 60^0$, and found that the
origin of the rotation was neither due to the emergence of new
flux, nor to the interaction with the surrounding field (the AR is
isolated). We found also a smaller fraction of ARs (5/22) that
rotate away from the East-West direction.  The number of cases in
each of the above categories is the same if we do not correct for
differential rotation.

\subsection{Twist versus writhe}
     \label{origin-twist-writhe}

    %PARAGRAPH Twist versus writhe plot: why doing this ? \\
     $\alpha_{\rm best}$ and $\Delta \varphi_{\rm cor.}$ are proxies
for the twist and for the writhe of the magnetic flux tubes,
respectively.  They are two independent measurements for each AR and
any correlation between them can give clues on the physical mechanisms
creating RTARs.

\begin{figure*}    %%%%%%%%%%%%%%%%%%%%%%%%%FIGURE 7
\centering
     \hspace{0cm}
\includegraphics[bb=54 360 558 720,width=17.cm]{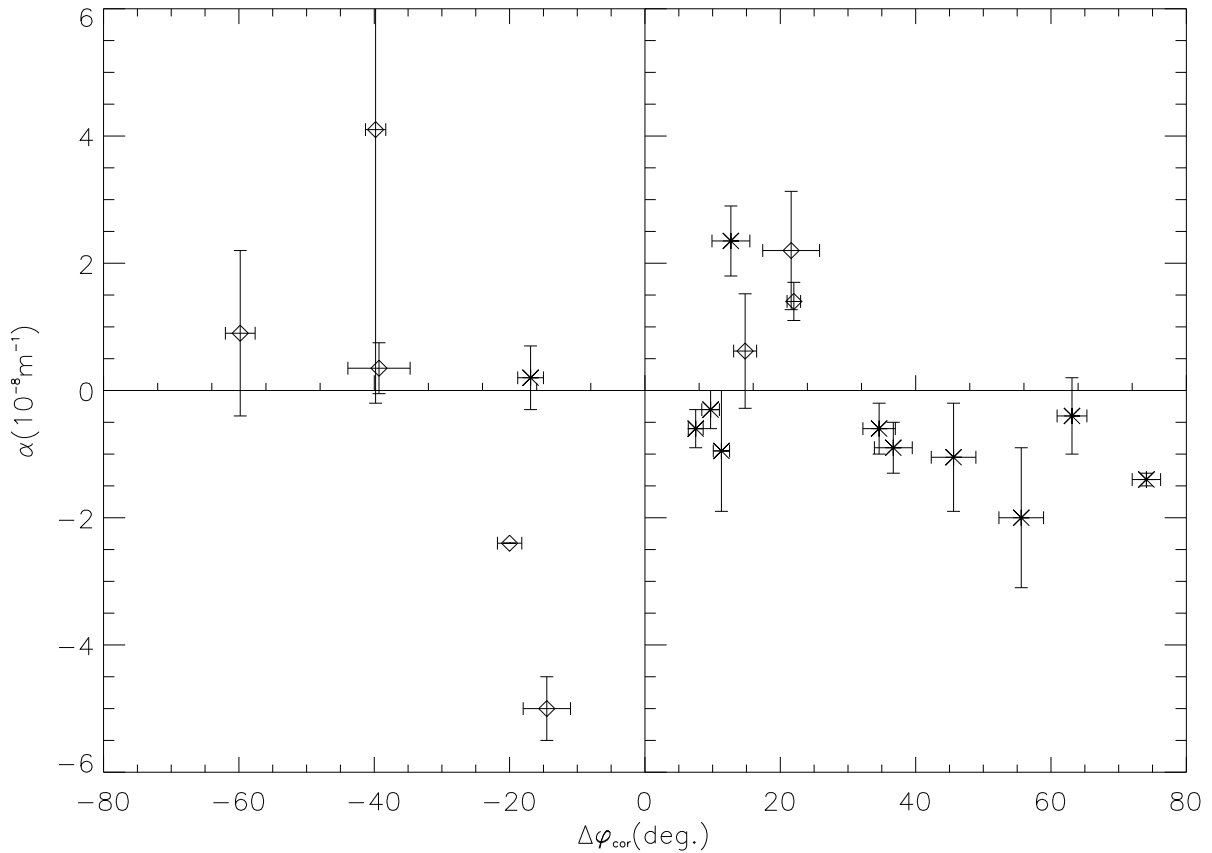}
         \caption{Twist vs writhe for the studied ARs.
The proxy for the twist is $\alpha_{\rm best}$ (Section~\ref{twist})
and the proxy for the writhe is total rotation angle corrected for
differential rotation (noted $\Delta \varphi_{\rm cor.}$, see
Section~\ref{writhe}). The value of $\alpha$ plotted in the ordinate
is the average of the measured values of $\alpha_{\rm best}$ for
each AR (see Tables~\ref{south} and~\ref{north}). North hemispheric 
RTARs are plotted with asteriks and South hemispheric ones 
with diamonds.  }
            \label{twist_writhe}
      \end{figure*}

    %PARAGRAPH Twist versus writhe plot: what is plotted \\
      Figure~\ref{twist_writhe} illustrates the results
shown in Tables~\ref{south} and~\ref{north}.  The values of
$\alpha_{\rm best}$ plotted in this figure
correspond to the average of the measurements for each AR.  We
have omitted AR 7882 for the sake of clarity, since both its
average $\alpha_{\rm best}$ and error bar are considerably larger
than the others. ARs 6855 and 8243 have also been omitted
since, as discussed in Section~\ref{twist-selectedARs}, the sign
of $\alpha_{\rm best}$ changes along their evolution. The
errors in $\alpha_{\rm best}$ correspond to the average of the
errors computed in the way described in Section~\ref{twist}. The
errors on $\Delta \varphi_{\rm cor.}$ are derived as explained in
Section~\ref{writhe-tilt-AR}.

    %PARAGRAPH Twist versus writhe plot: results\\
      In the set of 20 bipolar flux concentrations (ARs and their
reappearances), there are 7 that have the same sign of twist and
writhe and 13 for which the signs are different.  Some cases have
low values for $\alpha_{\rm best}$ or $\Delta\varphi$, so, one
wonders about their significance; however, taken as a whole
Figure~\ref{twist_writhe} indicates an anticorrelation between
twist and writhe.

\subsection{Comparison to previous studies}
     \label{origin-comparison}

    %PARAGRAPH Comparison to previous studies: generality\\
      Canfield and Pevtsov (\cite{Canfield98}) and 
Tian \etal (\cite{Tian01})
have analyzed the relationship between the tilt angle and the twist of
ARs.  Both studies differ from our analysis in two important points:
first, these authors analyze the ARs individually (they do not
identify all the reappearances of the same flux tube), and second,
they analyze the tilt (not the variation of the tilt).  In the
previous works, the tilt angle of each appearance, which is considered
as an independent region, is used as a proxy of the writhe of the
magnetic flux tube (we will refer below to this proxy as
writhe$_{(tilt)}$). This approach assumes that the flux tube is rooted 
to a
toroidal field, and that the observed tilt is a proxy for the
deformation of the flux tube from a planar geometry (for which the writhe
is null).
Then, the approach in Canfield and Pevtsov (\cite{Canfield98})
and Tian \etal (\cite{Tian01}) is related, but clearly different from the
one taken by us (a proxy of the writhe is the long-term evolution of
the flux concentrations with no hypothesis made about the rooted part
of the flux tube, see Section \ref{writhe-tilt} and Fig.~\ref{tubes}).

    %PARAGRAPH Comparison to previous studies\\
      The results of Canfield and Pevtsov (\cite{Canfield98}) show
a dominantly positive (resp.  negative) writhe$_{(tilt)}$ in the
northern (resp.  southern) hemisphere, which is a confirmation of
Joy's law.  They also find a dominant negative (resp.  positive) twist
in the northern (resp.  southern) hemisphere. But their results show
a positive correlation between the twist and writhe$_{(tilt)}$ (which
is not intuitive taking into account that both
hemispheric rules have been verified by them).
Tian et al.  (\cite{Tian01}) also find that
both hemispheric rules are fulfilled in their survey of 286 ARs, but
contrary to Canfield and Pevtsov, Tian et al.  show that 60~\% of the
ARs have opposite sign of twist and writhe$_{(tilt)}$ (in both
hemispheres).  Our results are similar to those of Tian et al., around
65~\% (13/20 cases for which the sign of the twist can be unambiguously
determined)
of the ARs show an anticorrelation between the signs
of writhe and twist.

\subsection{Conservation of magnetic helicity}
     \label{origin-helicity}

     The magnetic flux tubes that form ARs are most probably
formed at the base of the convective zone (CZ) from the global
toroidal component of the solar magnetic field (Parker
\cite{Parker93}).  Both 2-D (Emonet and Moreno-Insertis
\cite{Emonet98}) and 3-D models (Abbett \etal \cite{Abbett00}, Fan
\cite{Fan01}) predict that only flux tubes with a certain amount of
twist would survive to the interaction with the surrounding plasma as
they travel through the CZ.  Then, an AR observed at the photosphere
is formed by a flux tube which initially starts rising with some
finite amount of magnetic helicity in the form of twist ($T_{0}$).
Later on, during its transit through the CZ, its magnetic helicity
will remain roughly constant for the following reasons.
First, magnetic helicity is weakly dissipated in a plasma with a large
magnetic Reynolds number (it is dissipated only on the global
diffusion time scale, Berger \cite{Berger84}).  Second, the magnetic
helicity of the flux tube can change only if there is an input from
its footpoints (in the tachocline), but the dynamo works in a time
scale much longer than the CZ transit time of the tube (which typically
is of the order of one month).

     However, an internal transfer between twist and writhe helicity can
occur.  In particular, any helicoidal-like distortion of the flux tube
axis will introduce writhe helicity $W_{1}$.  Because of helicity
conservation, this induces a change in the twist helicity, which
becomes $T_{1}$, so that:
     \be
      T_{0}=T_{1}+W_{1}  \,. \label{EqHELI}
     \ee
There are various mechanisms that are able to transform the flux tube axis
to introduce writhe.  These can be: an internal instability of
the tube (such as the kink mode), or the Coriolis force acting on
the ascending tube, or the drag action of external convective
flows. Depending on the mechanism, $W_{1}$ is related in a different way to
some characteristic parameters of the flux tube
(such as $T_{0}$ or its latitude) so that the behavior of
$W_{1}$ traces the underlying mechanism.

      \begin{table}
         \caption[]{Results of the comparison between twist
and writhe.  We list the number
of ARs having the same (opposite) sign of twist ($T$) and writhe ($W$)
separately for the two hemispheres and added up.}
             \label{tablecompar}
       \(
              \begin{array}{ccc}
               \hline
               \noalign{\smallskip}
          Hemisphere & T.W > 0 & T.W < 0  \\

              \noalign{\smallskip}
               \hline
               \noalign{\smallskip}

               North & 2 & 10  \\
               South & 5 &  3  \\
               Total & 7 & 13  \\

               \noalign{\smallskip}
               \hline
            \end{array}
         \)
      \end{table}

\subsection{Kink instability}
     \label{origin-kink}

      %PARAGRAPH: summary of the kink instability \\
     The MHD kink instability has been studied by many authors in the
context of laboratory, coronal and CZ plasmas (see e.g.  Baty
\cite{Baty97}, Linton et al.  \cite{Linton99} and references there
in).  The instability occurs when a magnetic flux tube is twisted
above a critical value.  As the instability proceeds, the flux tube
untwists by one turn (for the most unstable mode $m=1$), reducing the
magnetic stress that originates the instability.  However, because of
the conservation of magnetic helicity (Eq.~\ref{EqHELI}), the twist
helicity is transformed into writhe helicity.  Then, in the new
equilibrium, the axis of the flux tube lies on a deformed helix with a
rotation of one turn.

      %PARAGRAPH: Why kink unstability for RTARs ?\\
     In contrast with a classical $\Omega$ loop, which is planar,
a flux tube where the kink instability has developed will have the
geometry shown in Fig.~\ref{tubes}.  The kink instability is then a
possible mechanism for the origin of the RTARs
(Section~\ref{writhe}).  However, one characteristic of the kink
instability is that $W_{1}$ should have the same handedness (i.e.  same
sign) as $T_{1}$ (because the instability develops only for large
twist so that $|T_{0}|>|W_{1}|$).

      %PARAGRAPH: summary of the results =>  not kink !\\
     The results of Section~\ref{origin-twist-writhe} show that only
7 (those marked with a $K$ on the right side of the last columns
in Tables~\ref{south} and~\ref{north}), over the 20 ARs for which
the sign of the twist can be determined, have the same twist
and writhe sign, but 13 ARs have opposite signs
(Table~\ref{tablecompar}). That is to say, at most only $\approx
35$~\% of the ARs can be associated with the formation of kinked
flux tubes (in the sense given by the kink instability). These ARs
tend in general to have large values of $\alpha_{best}$, as
illustrated in Figure~\ref{twist_writhe} (first and third quadrant); 
this agrees with the fact
that the kink instability develops in tubes having an excess of
twist. However, AR 7912 analyzed in L\'opez Fuentes 
\etal (\cite{Lopezf00}) is a clear example of an AR having a 
large twist, deduced from coronal field extrapolations (so, different 
from the approach used here), and still showing 
a different sign of twist and writhe.

      %PARAGRAPH  signs of twist and writhe equal does NOT imply kink ! \\
     It is worth mentioning that the fact that the signs of twist 
and writhe
are the same does not imply that the deformation of the flux tube is
due to the kink instability.  Having the same sign is a necessary, but
not a sufficient, condition to confirm the kink instability as the
process responsible of the deformation.  Other processes may still be
at the origin of the distortion of those tubes with the same sign of
twist and writhe.  Moreover, another physical process is necessarily
acting in ARs which have different sign of twist and writhe.

\subsection{Coriolis force}
     \label{origin-Coriolis}

      %PARAGRAPH: summary of the emergence of a flux tube \\
     A possible mechanism is the action of the Coriolis force on
the plasma during the emergence of the flux tube.  This has exhaustively been
studied in relation to Joy's law (D'Silva \&
Choudhuri \cite{Dsilva93}; Fan et al. \cite{Fan94}; Sch\"{u}ssler et
al. \cite{Schusler94}; Caligari et al. \cite{Caligari95}).  A review
on these works can be found in Fisher et al. (\cite{Fisher00}).

      %PARAGRAPH: summary of the Coriolis force effect \\
      As a flux tube emerges through the CZ, the Coriolis
force deforms its main axis introducing writhe helicity. If the
flux tube is in the North (South) hemisphere the
acquired writhe is positive (negative).  Because of
magnetic helicity conservation (Eq.~\ref{EqHELI}), this adds an
oppositely directed twist in the tube, which is negative (positive)
in the North (South) hemisphere.  This is in agreement
with the observed hemispheric rule for the mean twist
(Section~\ref{twist-hemispher}). In this scenario, the tilt angle
is expected, first to satisfy Joy's law on average (leader closer
to the equator); then, to change in the counterclockwise (clockwise)
direction in the northern (southern)
hemisphere.  If the flux tube remains anchored to the bottom of
the CZ through this long-term evolution, the expected final
direction of the bipole is East-West (the direction of the
toroidal field), while if disconnection occurs earlier this
direction will not be reached and the further evolution of the
magnetic field is the result of turbulent diffusion and advection
by large scale flows (Fan et al. \cite{Fan94}).

      %PARAGRAPH: RTARs rotating as Coriolis predicts\\
%     Do we have evidences for the above scenario~?
 To check the relevance of the Coriolis force as the main mechanism
to originate the flux tube deformation, let us first see how many
RTARs rotate in the right sense to be explained by this mechanism.
Looking at our results in Tables~\ref{south} and~\ref{north}, we
find that from the 13 ARs that cannot be associated with the kink
instability, 12 have the appropriate sense of rotation. Besides,
from the 7 ARs that can be related to a deformation of flux tubes
because of a kink instability, 3 can alternatively be originated
by the Coriolis force (they satisfy Joy's law and rotate towards
the East-West direction). These 3 cases should have an initial
twist $T_{0}$ that is opposite to the hemispheric rule (as
observed, see Tables~\ref{south} and~\ref{north}), of the same
sign and larger in magnitude than $W_{1}$, to still have $W_{1}$
and $T_{1}$ of the same sign.  Summarizing,
 %Next, if we include the 2 ARs for
 %which $\alpha_{best}$ changes of sign along their evolution (AR
 %6855 and AR 8243, see Tables~\ref{north} and~\ref{south}), 
we find that 15 out of 22 ARs rotate
in the direction expected when the Coriolis force has deformed the
flux tube.

  %PARAGRAPH: Adding relaxation, the non Coriolis cases\\
  If we also require that a flux tube deformed by 
the Coriolis force should evolve as described in the above second
paragraph, relaxing to the East-West direction (see also
Section~\ref{relaxation}), we find that only 9 of the 15 ARs that
rotate in agreement with the action of Coriolis force are coherent with
that scenario (those marked with a $C$ on the right side
of the last columns in Tables~\ref{south} and~\ref{north}). 
Therefore, only $\approx$ 41\% of the full set of
studied RTARs (9/22 cases) fulfill these two requirements (correct
sense of rotation for the Coriolis force and relaxation to the
toroidal direction). We conclude that, under our assumptions, the
rotation of the polarities for most of the studied ARs is not
coherent with the effect of the Coriolis force on the ascending
flux tube.

\subsection{Convective-zone motions}
     \label{origin-convective}
    % generality on the possible mechanism \\
     Section~\ref{origin-kink} shows that the dominant mechanism driving
the variation of the tilt angle does not have an internal origin
in the magnetic flux tube. We have analyzed only the kink
mode, since buoyancy and pressure are
unable to distort the flux tube axis on an helix-like path. 
We have also shown that the Coriolis force can only account for 
9 out of 22 of the studied cases.
The RTARs, which can be attributed neither to the kink
instability nor to the action of the Coriolis force, have been
marked with a $?$ (for unidentified) on the left side of the last
columns in Tables~\ref{south} and~\ref{north}, and they also amount
to $\approx$ 41\% of the full set (9/22 cases). At this point, we
have to consider the interaction of the flux tube with the
external plasma via the drag force, which can be very efficient in
coupling them, specially if the subphotospheric flux tube is split
in several thinner tubes (e.g. Zwaan \cite{Zwaan87}).

    % More precise observation and theory \\
      Large scale vortex motions are presently difficult to infer from
the data. Nevertheless, Ambro{\v z} (\cite{Ambroz01b} and
references therein) claimed for the detection of large scale
vortex with a spatial size of the order of 200 Mm and with a time
scale of 4 Carrington rotations. The results are deduced using the
Wilcox Solar Observatory synoptic maps and a local correlation
tracking algorithm. Such vortex motions, which may be present
at the photosphere or at larger depths, are still an open possibility
to explain the origin of some RTARs. They may act deforming a flux tube
from its planar $\Omega$ shape while it travels through the CZ,
or after the emergence of a classical planar $\Omega$-loop forcing
the photospheric polarities to rotate. On the 
theoretical side, Longcope \etal (\cite{Longcope98}) propose that the 
turbulent velocities in the CZ can, via the drag force, 
deform the axis of an ascending
magnetic flux tube. The writhe of the flux tube is modified
inducing twist via magnetic helicity conservation
(Eq.~\ref{EqHELI}). The derived mean tilt angle and its dispersion
compare favorably with the corresponding observations. Although
the buffeting of flux tubes by CZ turbulence can be invoked as the
origin of the variation of the tilt angle found in the 22 studied
ARs, our selection criteria (clear rotation and availability of
vector magnetograms) compel us to use a too biased and too small
sample, and do not let us arrive at a conclusion about a
mechanism which is by nature of statistical origin.

%%%%%%%%%%%%%%%%%%%%%%%%%%%%%%%%%%%%%%%%%%%%%%%%%%%%%%%%%%%%%%%%%%%%
\section{Conclusion}
\label{conclusion}

     %PARAGRAPH Aims of this study \\
      We focus this study on the long term evolution of a set of
bipolar active regions (ARs) in which the main polarities were
seen to rotate one around the other (named as rotating tilt-angle
ARs, RTARs). 
%We interpret this peculiar evolution as being the
%result of the emergence of magnetic flux tubes, which are
%distorted with respect to the classical planar $\Omega$-loop
%shape.  
Differential rotation contributes only partially to the
change of the tilt angle of the bipoles and it is not the main
mechanism. In principle, a possible origin of the observed
evolution is the nonlinear development of a kink-instability that
occurs in the convection zone (CZ) when the tube is twisted above
a critical value. The development of this instability creates a 
flux tube with a non-planar axis geometry (introducing magnetic
helicity in the form of writhe). Then, RTARs can result from the 
emergence of magnetic flux tubes that are distorted with respect to 
the classical planar $\Omega$-loop shape. 
The main objective of this study has been to
test this possibility using the largest available set of ARs.  
We have inferred the sign of the writhe of RTARs computing the 
variation of their tilt angle from synoptic maps and we have used 
photospheric vector magnetograms to determine the sign of the magnetic 
twist. The intersection of the set of RTARs and the set of ARs for
which vector magnetograms are available amounts to 22 cases.

     %PARAGRAPH Summary of the results for the kink\\
     A main characteristic of the kink instability is that the
flux tube axis is distorted having the same handedness as the
twist, so this mechanism necessarily implies that the sign of
twist and writhe should be the same.  Comparing the handedness of
the magnetic twist and writhe, we find that the presence of
kink-unstable flux tubes is coherent with no more than 35\% of the
studied cases; so, at most, only a fraction (7/20) of these RTARs
can be explained by this process. This confirms previous results
derived when investigating a few cases (only 2 cases over 5 could
be explained by the kink instability, see L\'opez Fuentes et al.
\cite{Lopezf00}). With such a small percentage, we conclude that
the kink instability cannot be the main mechanism at the origin of
the observed rotation of the polarities in the studied subset of
ARs.

     %PARAGRAPH Other mechanism ?\\
     Another possible mechanism is the action of the Coriolis force on
the ascending flux tubes in the CZ.  In particular, Coriolis force
is invoked to explain Joy's law (leader spot closer to the
equator). However, we have found no systematic relaxation of the
bipoles towards the toroidal direction, rather an important
fraction of the ARs (7/22) rotates towards the toroidal direction
and then overpasses it by more than $10^{0}$. A similar fraction
(5/22) rotates away from the toroidal direction. Therefore, only
$\approx$ 41\% (9/22) of the RTARs can be originated by the
Coriolis force. Another possible mechanism, which can create 
the RTARs (9/22) that cannot be explained either by the kink 
instability, or by the Coriolis force, is the action of large
scale vortexes and/or turbulent flows in the CZ or photosphere; 
these may couple to the ascending or already emerged flux tube through 
the drag force.

     Summarizing, after showing that differential rotation cannot
explain the change of the tilt, our results demonstrate that 
none of the two mechanisms we
have tested can be considered as the main cause of the
deformation of the flux tubes forming the RTARs. The fraction of ARs 
that can be explained by any of them is equivalent: 7/20 for the kink 
instabilty and 9/22 for the Coriolis force. 
Furthermore, none of the mechanisms excludes 
the other, which makes it even more difficult to distinguish between them.
For the cases that can be attributed to neither of them (9/22), we 
have proposed the action of large scale vortexes in the CZ, 
or in the photosphere or near sub-photosphere. These 
motions impose no clear signature on the emerging flux tubes and,
in fact, they could be acting on all the RTARs.
Moreover, if RTARs rotate mainly because of photospheric or
shallow sub-photospheric motions; then, any signature of the processes
that could deform the flux tube before emergence might be completely
washed out.      
As mentioned above, our sample is too biased by the kind of ARs 
that we have chosen to analyze, those showing a coherent change in 
their tilt 
angle along several solar rotations. Besides, the fact that we wanted to
test the relevance of the kink instability restricted our sample even
more.  
It will be the objective of a next paper to study the
kinetic properties of a much larger set of ARs (L\'opez Fuentes et
al., in preparation) to determine the role of the Coriolis force and of
large scale turbulent plasma motions in the CZ on the observed large
tilt angle variation.

     %PARAGRAPH Kink instability: delta spot; need a close check ! \\
     Finally, it is worth mentioning that models of kink-unstable tubes have
been proposed to explain a particular kind of ARs, the so-called
$\delta$-spot configurations (Linton \etal \cite{Linton98},
\cite{Linton99}; Fan \etal \cite{Fan99}).  Basically, they are
formed by two strong opposite polarities very close together
sharing  the same penumbra.  Such configurations, if present in
ARs, appear in their early stages, while our study addresses the
long-term evolution of ARs.  Then, it has to be tested on a daily
basis, whether $\delta$-spot configurations rotate in such a way
that their writhe and twist have dominantly the same sign. This
first test will tell us if $\delta$-spot configurations are likely
to be a consequence of the kink-instability in the CZ.  If this
first test is successful, a further test would be to estimate the
amount of twist in the flux tube to check if it is large enough to
trigger the kink instability (this critical value depends on the
twist distribution in the tube). Probably, this second test will
be possible in a near future with the next generation of vector
magnetographs, and the computation of the coronal magnetic
helicity budget (a way to estimate the total helicity present
initially in the flux tube, see D\'emoulin et al.
\cite{Demoulin02a}).

\begin{acknowledgements}
      We acknowledge the referee, Dr. E. N. Parker, for his 
helpful suggestions.
      M.L.F. acknowledges the Observatoire de Paris, Meudon, for
financial support.
C.H.M. and P.D. acknowledge financial support from ECOS (France) and SETCIP
(Argentina) through their cooperative science program (A01U04).
      L.v.D.G.was supported by the Research Fellowship F/01/004 of the
K.U. Leuven and by the Hungarian Government grants OTKA
T-032846, T-038013. P.D. and L.v.D.G. acknowledge the Hungarian-French S\&T
cooperative program.
      The NSO/Kitt Peak data used here are produced cooperatively by NSF/NSO,
NASA/GSFC, and NOAA/SEL. The National Solar Observatory (NSO) is operated
by the Association of Universities for Research in Astronomy
(AURA, Inc) under cooperative agreement with the National Science
Foundation (NSF).

\end{acknowledgements}

\end{document}